# MRI-only brain radiotherapy: assessing the dosimetric accuracy of synthetic CT images generated using a deep learning approach

Samaneh Kazemifar, Sarah McGuire, Robert Timmerman, Zabi Wardak, Dan Nguyen, Yang Park, Steve Jiang, Amir Owrangi

Medical Artificial Intelligence and Automation Laboratory, Department of Radiation Oncology, University of Texas Southwestern, Dallas, Texas, US

Emails: Amir.Owrangi@UTSouthwestern.edu

**Abstract:**

**Purpose:** This study assessed the dosimetric accuracy of synthetic CT images generated from magnetic resonance imaging (MRI) data for focal brain radiation therapy, using a deep learning approach.

**Material and Methods:** We conducted a study in 77 patients with brain tumors who had undergone both MRI and computed tomography (CT) imaging as part of their simulation for external beam treatment planning. We designed a generative adversarial network (GAN) to generate synthetic CT images from MRI images. We used Mutual Information (MI) as the loss function in the generator to overcome the misalignment between MRI and CT images (unregistered data). The model was trained using all MRI slices with corresponding CT slices from each training subject's MRI/CT pair.

**Results:** The proposed GAN method produced an average mean absolute error (MAE) of 47.2 ±11.0 HU over 5-fold cross validation. The overall mean Dice similarity coefficient between CT and synthetic CT images was 80% ± 6% in bone for all test data. Though training a GAN model may take several hours, the model only needs to be trained once. Generating a complete synthetic CT volume for each new patient MRI volume using a trained GAN model took only one second.

**Conclusions:** The GAN model we developed produced highly accurate synthetic CT images from conventional, single-sequence MRI images in seconds. Our proposed method has strong potential to perform well in a clinical workflow for MRI-only brain treatment planning.

**Introduction:**

Treatment planning in modern radiation therapy uses both computed tomography (CT) and magnetic resonance imaging (MRI) for many disease sites. Although CT images provide electron density values needed for treatment planning, MRI images provide superior soft tissue contrast to delineate tumors and soft tissues. Moreover, incorporating MRI into treatment planning substantially reduces inter- and intra-observer contouring variability for many disease sites [1-4]. In the brain, MRI can resolve tumor boundaries that cannot be resolved on CT [5] and can identify peritumoral edema [6]. Furthermore, MRI is a multi-parametric imaging modality that can provide not only anatomical information with high soft-tissue contrast, but also valuable functional information for assessing disease progression and evaluating treatment response [7, 8].

Treatment planning based only on MRI would reduce radiation dose, patient time, and imaging costs associated with CT imaging, streamlining clinical efficiency and allowing high-precision treatment planning. Additionally, anatomic and functional imaging needed for treatment planning could all be performed during the same sophisticated imaging session, thereby reducing image registration errors. Despite these advantages, several challenges must be overcome before MRI-only planning can be introduced into the clinic, including developing robust methods to accurately generate synthetic CT images from MRI images.

The methods proposed in the literature to automatically generate synthetic CT images from MRI images can be divided into different categories, including tissue segmentation-based, learning-based, atlas-based, and deep learning-based approaches. For example, Paradis *et al.* [9] applied several pulse sequences (including UTE) using a voxel-based technique to generate synthetic CT brain images, and they achieved a mean dose difference in the target of less than 1%. Other groups used T1/T2* MRI images to generate synthetic CT images in brain or prostate using voxel-based methods. Several studies used atlas/multi-atlas–based methods for this purpose with a mean absolute error (MAE) of 184 ± 34 in brain [10] and a range of 36.5 ± 4.1 and 54 HU in prostate [11, 12].

Han [13] was the first to apply a 2D U-Net model with MAE as the loss function to convert a 2D MRI slice to its corresponding 2D CT of brain. They reported an average MAE of 84.8 ± 17.3 across all test data and required an accurate MRI/CT alignment in the training data. Maspero *et al.* [14] used 3 contrast MRI images, including in phase, water, and fat images, as input for a generative adversarial network (GAN) in pelvis data with an average MAE of 61.0 ± 9.0 over all test data. Also, they rigidly aligned and assigned air locations from MRI to CT. GAN has been used to generate synthetic CT from unpaired brain MRI images [15] and paired training data sets [16]. Although deep learning-based methods have achieved state-of-the-art performance across different medical applications [17], they can still be improved for optimal clinical and research applications [18]. So far, no study has evaluated the performance of a GAN model with a new loss function or evaluated the dose calculation accuracy of synthetic CT generated using this modified model on brain MRI. Therefore, we investigated the accuracy of the proposed method and performed a dosimetric evaluation of synthetic CT images generated from un-aligned MRI/CT data for focal brain radiation therapy using a deep learning approach.

## Material and Methods:

**Image acquisition**

We analyzed CT and MRI images from patients who had undergone brain tumor radiotherapy. Tumor sizes varied between 1.1 and 42.4 cm$^3$. The images were collected at the University of Texas Southwestern Medical Center (UTSW) as part of the standard treatment protocol. CT and MRI images of 77 patients were retrospectively reviewed using an institutional cancer registry database under an institutional review board-approved study (IRB number: STU 052012-019). Patients underwent both CT (Philips Big Bore scanner, Royal Philips Electronics, Eindhoven, The Netherlands) and MRI scanning for radiotherapy treatment planning. Because this is a retrospective study and the CT and MRI scans were performed in different departments, images were acquired using different vendors. All CT images were acquired in the Department of Radiation Oncology using a 16-slice CT, 120 kV, exposure time= 900 ms and 180 mA. Images were acquired with a 512 x 512 matrix and 1.5 mm slice thickness (voxel size $0.68 \times 0.68 \times 1.50$ mm$^3$). All MRI images were acquired in the Department of Radiology using a 1.5T magnetic field strength and a post-gadolinium 2D T$_1$-weighted spin echo sequence with TE/TR = 15/3500 ms, a 512 x 512 matrix and an average voxel size of $0.65 \times 0.65 \times 1.5$ mm$^3$. The pixel size range of MRI data was 0.51 x 0.51mm$^2$ - 0.88 x 0.88mm$^2$. To train the model, the MRI images were resampled to the same voxel size as CT images. In addition, the synthetic CT have the same voxel value as CT images (0.68 x 0.68 x 1.5 mm$^3$). 70% and 12% of the patient data was randomly selected for training and validation, respectively, and the remaining 18% was used for testing. We used 5-fold cross validation to evaluate the model's performance. 82% of the data was split into 5 folds, and in each run, 4 folds were used for training and one fold was used for validation. Each trained model was used to generate synthetic CT images using the test data. Test data were never used for training and validation; they were only used at the end to evaluate the accuracy of this method.

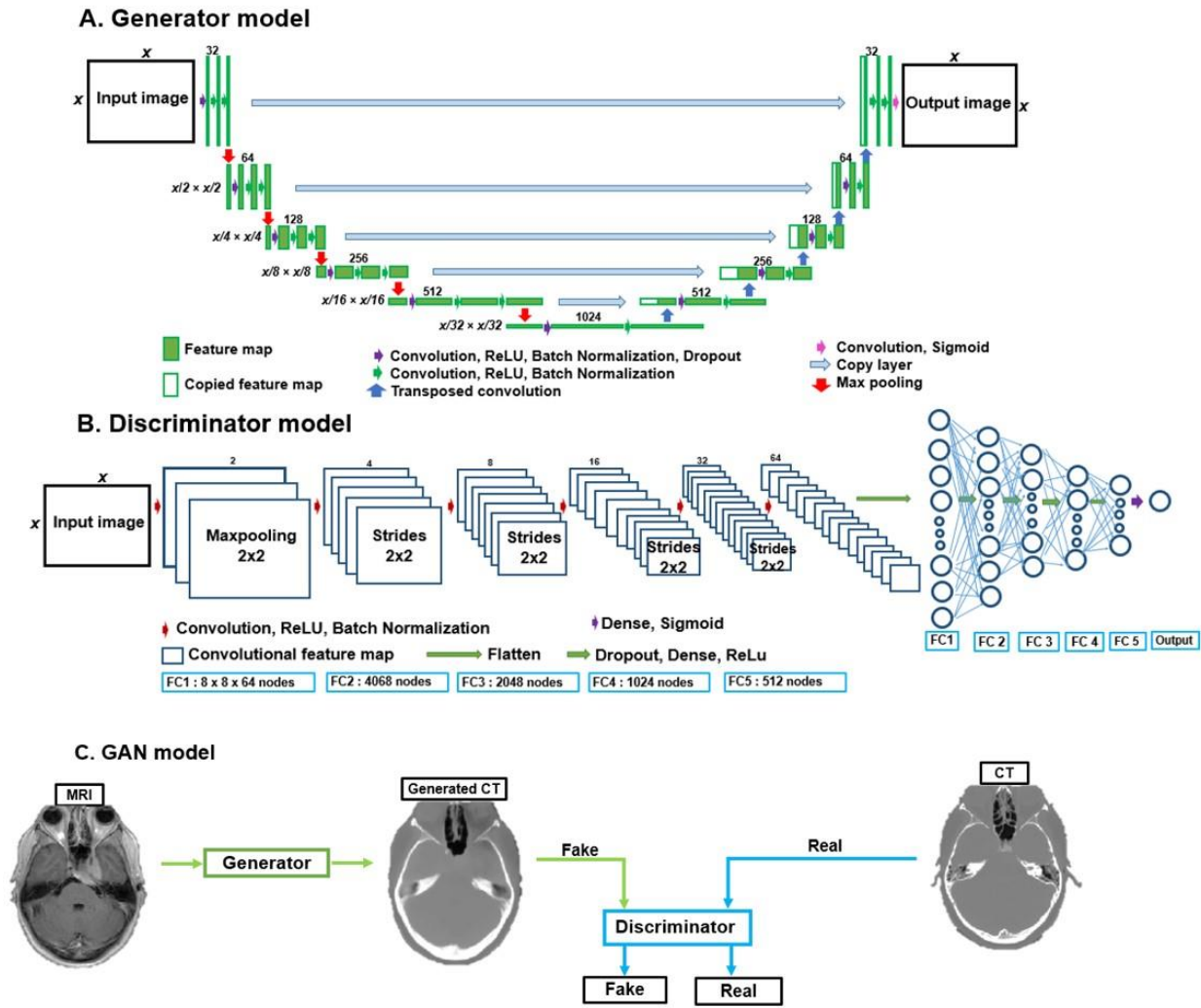

**Figure 1.** Proposed GAN model with two networks, such as generator and discriminator models. The structures of the Generator (A) and Discriminator (B) are shown. GAN model for generating synthetic CT from MRI images (C).

**Generative adversarial network (GAN):**

Generative adversarial networks [19] are a class of deep machine learning algorithms used in unsupervised learning and implemented by two convolutional neural networks: a generator and a discriminator. Deep learning methods are the current state-of-the-art for generating HU maps because they have the advantage of learning low features to complex features automatically from images. The most appropriate networks for generating synthetic CT from MRI images were identified by Ronneberger et al. as the U-Net [5] and by Goodfellow et al. as the generative adversarial network (GAN) [6]. Here, we used the Mutual Information (MI) as the loss function, which differs from the loss function used in a standard conditional GAN model [20]. For the generator network, we used a 2D U-Net [21] model that directly learns a mapping function to convert a 2D MRI grayscale image to its corresponding 2D synthetic CT image. Generally, the U-Net model is a pixel-wise prediction network and includes convolutional (encoder) and deconvolutional (decoder) networks. First, the low level feature maps are down-sampled to high level feature maps using a max pooling layer. Second, the high level feature maps are up-sampled to low level feature maps using the transposed convolutional layer to construct the prediction image. Our network contains blocks of convolutional 2D layers with variable filter sizes but the same kernel sizes and activation functions, except the last layer. The structure of our U-Net model is illustrated in Figure 1 (A). On the left side of the U-Net structure, we used three 3x3 convolutional layers, each followed by a rectified linear unit (ReLU) (activation function), and one max pooling operation. On the right side of the U-Net structure, we used a 2x2 transposed convolutional layer followed by a concatenate layer and two 3x3 convolutional layers with a ReLU activation function. In addition, we added a batch normalization layer to each 3x3 convolutional layer and a dropout layer to one 3x3 convolutional layer. For the final layer, we used a 1x1 convolutional layer with a filter size of 1 and a sigmoid activation function. The generator's loss function was MI using an Adam optimizer of learning rate=0.0002, beta_1=0.5. The discriminator takes an input with an image size of 352x352 and produces output as a real number in [0; 1]. The discriminator consists of 6 convolutional layers with different filter sizes but the same kernel sizes and strides, followed by 5 fully connected layers. We used ReLU as the activation function and a batch normalization layer for the convolutional layers. The dropout layer was added to the fully connected layers, and a sigmoid activation function was used in the last fully connected layer. We employed binary cross entropy for the loss function and the Adam optimizer with a learning rate=0.00005, beta_1=0.5 in this network. We used 3x3 for kernel size and 2, 4, 8, 16, 32 and 64 as the filter sizes in the discriminator network. The details of this structure are shown in Figure 1 (B). The interaction between the generator and discriminator networks in our GAN model is shown in Figure 1 (C). This work applied the concept of conditional GAN [20], but there are some differences from the original. First, we used U-Net with MI as the loss function, and second, we used several convolutional layers and several fully connected layers with ReLU and binary cross entropy as the activation and loss functions in the discriminator network.

We replaced the loss function of the generator model with MI between the CT and generated CT images, because MRI and CT images are not aligned. We formulated the MI inside the loss function to overcome difficulties in ideal registration. Mutual Information measures the "amount of information" of one variable when another variable is known. Maximizing Mutual Information is equivalent to minimizing the joint entropy (joint histogram). The MI between two variables is expressed as:

$$MI\left(x_i, G(y_i)\right) = \sum_{x_i, G(y_i)} p(x_i, G(y_i)) \log \frac{p(x_i, G(y_i))}{p(x_i) p(G(y_i))} = H(x_i) + H\left(G(y_i)\right) - H(x_i, G(y_i))$$

where $p(x_i, G(y_i))$ is the joint distribution, and $p(x_i)$ and $p(G(y_i))$ indicate the distribution of images $x_i$ and $G(y_i)$. Here, the loss functions are explained. Step 1 involves updating the discriminator D by the following loss function:

$$-\log D(x_i) - \log\left(1 - D\left(G(y_i)\right)\right) \qquad Eq.1$$

and Step 2 involves updating the generator G by the following cost function:

$$MI(x_i, G(y_i)) \qquad Eq.2$$

where $G$ is the generator and $D$ is the discriminator, $\{x_i, y_i\}$ is the training pair, $x_i$ is the CT image, $G(y_i)$ is the generated CT image with $y_i$ as the MRI image, $i$ is number of the image, *MI* is Mutual Information, H($x_i$) is the entropy of image $x_i$, and H($x_i$, $G(y_i)$) is the joint entropy of these two images.

**Table 1:** Mean difference percentages (range) between CT and synthetic CT across all test data with respect to $D_{max}$, $D_{mean}$, $D_{95\%}$, and $D_{5\%}$ in target and critical structures. NS: not significant, paired two-tailed *t*-test. The third column shows results using the same plan for CT and synthetic CT. The fifth column shows results using separate optimized plans for CT and synthetic CT.

|  |  | CT vs synthetic CT mean differences (range) | *p*-value | CT (re-optimized plan) vs synthetic CT mean differences (range) | *p*-value |
|---|---|---|---|---|---|
| **PTV** | | | | | |
| | $D_{mean}$ (%) | -0.7 (-1.8 to 1.2) | NS | -0.9 (-2.8 to 1.6) | NS |
| | $D_{max}$ (%) | -0.6 (-2.3 to 0.3) | NS | -0.8 (-2.6 to 1.3) | NS |
| | $D_{95\%}$ (%) | -0.6 (-2.2 to 0.3) | NS | -0.7 (-2.5 to 0.9) | NS |
| | $D_{5\%}$ (%) | -0.7 (-2.2 to 0.4) | NS | -0.9 (-2.3 to 0.8) | NS |
| **Brainstem** | | | | | |
| | $D_{mean}$ (%) | -0.3 (-2.4 to 5.3) | NS | -0.6 (-2.8 to 5.7) | NS |
| | $D_{max}$ (%) | -0.8 (-2.4 to 0.4) | NS | -0.5 (-2.9 to 1.4) | NS |
| | $D_{95\%}$ (%) | 0.3 (-3.8 to 3.0) | NS | 0.9 (-2.8 to 2.0) | NS |
| | $D_{5\%}$ (%) | -0.7 (-2.5 to 0.7) | NS | -0.6 (-2.8 to 0.5) | NS |
| **Left Eye** | | | | | |
| | $D_{mean}$ (%) | -0.4 (-1.9 to 2.6) | NS | -0.7 (-1.6 to 2.3) | NS |
| | $D_{max}$ (%) | -0.4 (-1.5 to 0.5) | NS | -0.6 (-1.1 to 0.7) | NS |
| | $D_{95\%}$ (%) | -0.8 (-2.9 to 1.3) | NS | -0.5 (-2.2 to 1.6) | NS |
| | $D_{5\%}$ (%) | -0.4 (-1.9 to 2.6) | NS | -0.6 (-1.3 to 2.1) | NS |
| **Right Eye** | | | | | |
| | $D_{mean}$ (%) | -0.6 (-2.3 to 0.4) | NS | -0.9 (-2.7 to 1.4) | NS |
| | $D_{max}$ (%) | -0.5 (-2.3 to 0.9) | NS | -0.8 (-2.6 to 0.6) | NS |
| | $D_{95\%}$ (%) | -0.7 (-3.3 to 0.1) | NS | -0.5 (-2.8 to 1.1) | NS |
| | $D_{5\%}$ (%) | -0.4 (-2.1 to 1.3) | NS | -0.7 (-2.6 to 1.1) | NS |
| **Left Optic Nerve** | | | | | |
| | $D_{mean}$ (%) | -0.2 (-2.7 to 2.9) | NS | -0.6 (-2.3 to 2.3) | NS |
| | $D_{max}$ (%) | -0.1 (-2.1 to 2.2) | NS | -0.5 (-2.7 to 2.6) | NS |
| | $D_{95\%}$ (%) | -0.6 (-2.0 to 0.8) | NS | -0.4 (-2.3 to 1.8) | NS |
| | $D_{5\%}$ (%) | -0.4 (-1.5 to 0.2) | NS | -0.6 (-1.1 to 1.2) | NS |
| **Right Optic Nerve** | | | | | |
| | $D_{mean}$ (%) | 0.1 (-0.4 to 1.0) | NS | 0.3 (-0.6 to 1.1) | NS |
| | $D_{max}$ (%) | 0.2 (-0.7 to 2.2) | NS | 0.4 (-0.5 to 2.0) | NS |
| | $D_{95\%}$ (%) | -0.4 (-1.7 to 0.5) | NS | -0.3 (-1.3 to 0.9) | NS |
| | $D_{5\%}$ (%) | 0.0 (-0.6 to 2.4) | NS | 0.1 (-0.5 to 1.4) | NS |
| **Optic Chiasm** | | | | | |
| | $D_{mean}$ (%) | -0.4 (-1.9 to 1.4) | NS | -0.5 (-1.3 to 1.2) | NS |
| | $D_{max}$ (%) | -0.4 (-2.4 to 1.5) | NS | -0.6 (-2.2 to 1.1) | NS |
| | $D_{95\%}$ (%) | -0.5 (-2.4 to 1.5) | NS | -0.3 (-1.7 to 1.6) | NS |
| | $D_{5\%}$ (%) | 0.0 (-2.4 to 2.0) | NS | 0.5 (-1.4 to 2.5) | NS |
| **Total MU (%)** | | 0.5 (-1.9 to 2.1) | NS | 0.6 (-1.7 to 1.6) | NS |

**Volumetric modulation arc therapy (VMAT) planning and dose calculation**

Both images were imported into the Eclipse treatment planning system (Eclipse v15.0, Varian Medical Systems). Then, the synthetic CT images were rigidly aligned to the CT images for dose comparison using Eclipse software. Targets and organs at risk (OARs)—including left and right eye, optic nerves, optic chiasm, and brainstem—were contoured on CT images and reviewed by the radiation oncologist. In the next step, CT contours were transferred to the registered synthetic CT images, and the optimized VMAT plan was transferred from CT to synthetic CT images. The dose was calculated on synthetic CT images using the plan transferred from the original CT images. In addition, we optimized the plans for CT and synthetic CT separately and measured mean dose values in planning target volumes (PTV) and OARs.

Dose volume histogram (DVH) was analyzed for all test patients, and dosimetric parameters and total monitor units (MUs) were compared between real CT and synthetic CT images. Finally, 3-D gamma analysis of dose distributions was performed at both 2%/2mm and 1%/1mm dose difference and distance levels.

**Results:**

**Comparison between synthetic CT and real CT**

To evaluate the accuracy of synthetic CT images generated for treatment planning, we calculated a mean absolute error (MAE) in HU for the whole external body contour for each patient. The comparison between real CT and synthetic CT and the difference map for axial, coronal, and sagittal views using the proposed method are shown for one representative patient in Figure 2. The results of the cross validation using the MI for all test data are shown in Figure 3. The average MAE ± SD value between CT and synthetic CT images for all test data using training sets 1, 2, 3, 4, and 5 were 41.8 ± 10.0, 48.8 ± 13.0, 48.2 ± 12.4, 48.2 ± 12.2, and 48.3 ± 12.0, respectively. The average of the average MAE over all cross validation sets was 47.2 ± 11.0. We also ran the model with MAE as the loss function instead of MI. The average MAE ± SD value was 60.2 ± 22.0, which was significantly larger ($p$-value< 0.05; paired two-tailed $t$-test) than our proposed loss function. In addition, the Dice similarity coefficient (DSC) between CT and synthetic CT images was 80% ± 6% in bone and 70% ± 7% in air. We used a ≥ 200 HU threshold for bone segmentation and a < -200 HU threshold for air segmentation. The mean DSC value between whole head CT contours and whole head synthetic CT contours was 96% ± 2% over all test patients with a range of 91% to 98%.

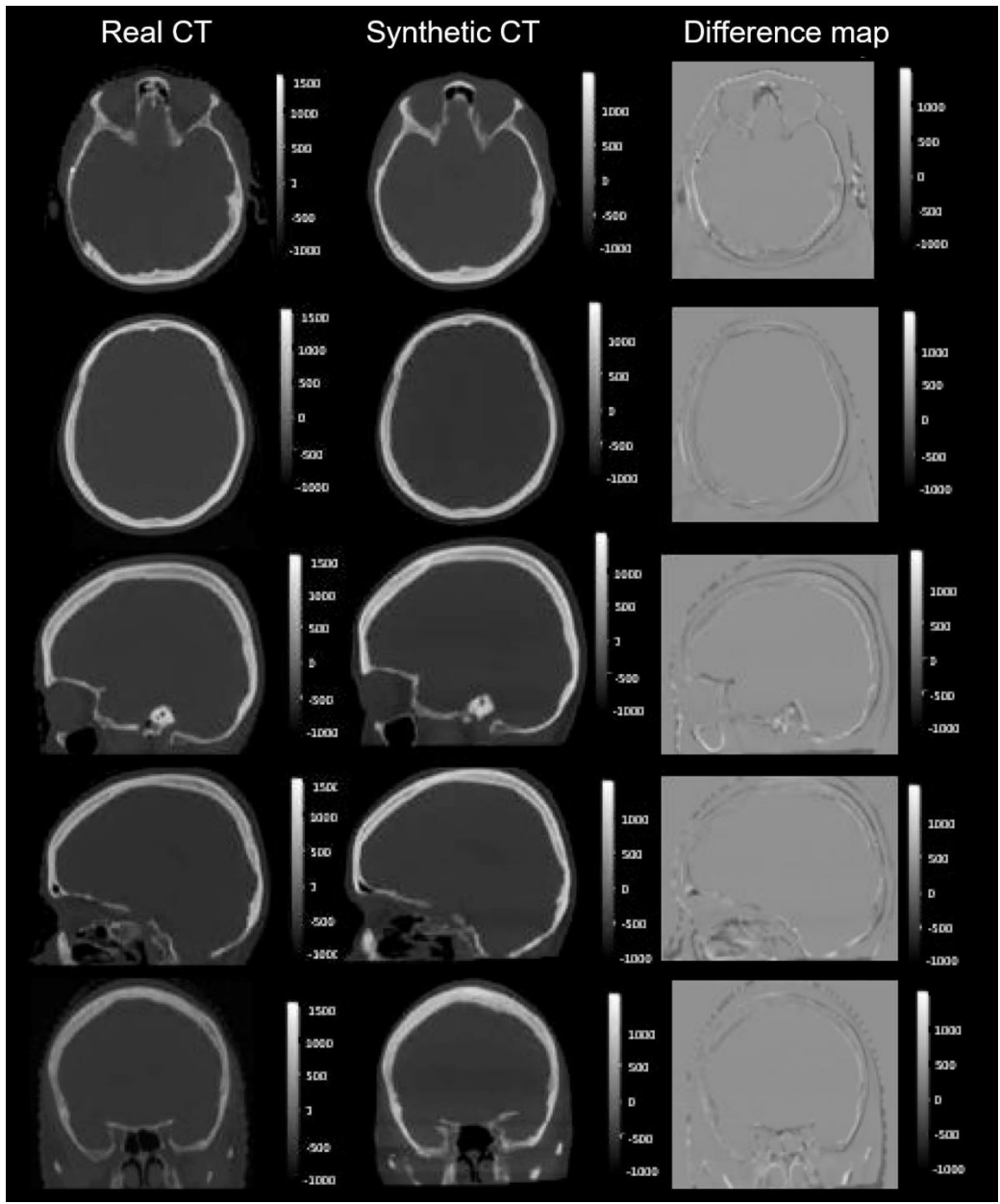

**Figure 2.** Comparison between real CT and synthetic CT from a patient's MRI brain image in the axial, sagittal and coronal views.

**Comparison between CT and synthetic CT doses**

For each patient, a VMAT plan was generated and the DVH was analyzed for target and critical structures. DVH parameters—mean dose ($D_{mean}$), minimum dose ($D_{min}$), maximum dose ($D_{max}$), $D_{95\%}$, and $D_{5\%}$—were calculated for PTV, brainstem, left/right eye, left/right optic nerve, and optic chiasm. $D_{95\%}$ and $D_{5\%}$ were defined as the minimum dose delivered to 95% and 5%, respectively, of the PTV. Table 1 shows the mean and range of percent differences in PTV and critical structures across all 14 patients using the same plan for CT and synthetic CT and also separate plans optimized for CT and synthetic CT. The mean percent difference between the doses calculated in CT and synthetic CT images was statistically insignificant and less than 1% overall for all DVH parameters. The mean percent difference between the MUs calculated in plans generated using CT and synthetic CT images was less than 1% and statistically insignificant. The calculated dose map for a single CT plan and synthetic CT plan from the same patient is shown in Figure 4 (A). The average DVH plots for all 14 patients are reported in Figure 4 (B). Gamma analysis between synthetic CT and CT plans using 2%/2mm and 1%/1mm acceptance criteria revealed mean passing rates of 98.7% and 93.6%, respectively, for 2D gamma analysis, and 99.2% and 94.6%, respectively, for 3D gamma analysis. Individual and mean passing rates of the two criteria are summarized in Table 2. A representative sample of the 3D and 2D (1%/1mm) gamma map for one patient is shown in Figure 5.

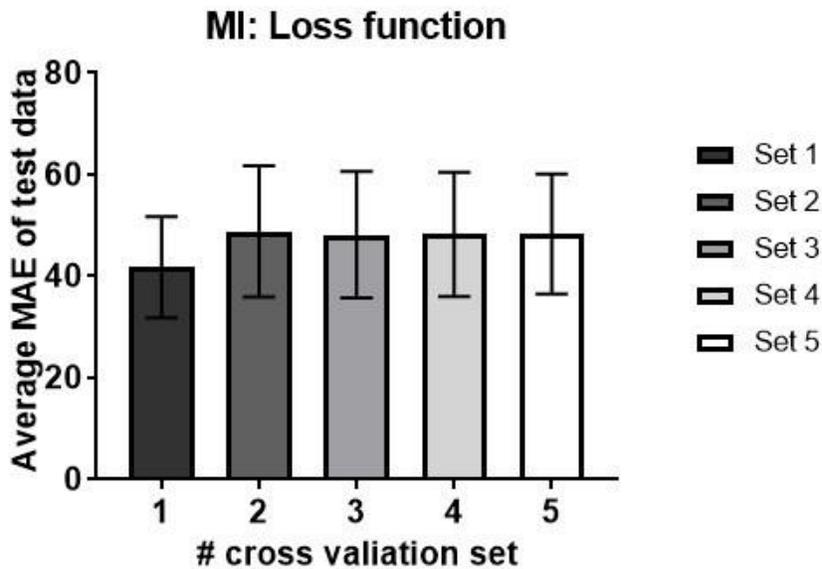

**Figure 3.** An average MAE ± SD over all test data (14 patients) using models trained on different training and validation data sets (Set 1, Set 2, Set 3, Set 4, Set 5). Mutual Information (MI) was used as the loss function in the generator network.

**Discussion:**

We investigated a new deep learning method to generate brain synthetic CT images that only requires one MRI pulse sequence to accurately display all regions, in contrast with classical image processing techniques that require multiple MRI modalities acquired with different pulse sequences. Most importantly, our method does not need pulse sequences such as UTE to accurately display bone-air interfaces. To minimize the alignment error of the unpaired MRI and CT images, we used MI between CT and generated CT images as the loss function in the generator section of the GAN. We evaluated the synthetic CT images generated by comparing their intensity using MAE, and results showed high agreement between the HU maps of both images. We also investigated the accuracy of the generated HU maps by comparing the dose distributions generated in both image sets. DVH and gamma analysis showed high agreement between both dose distribution sets. To measure a geometric score, we used the DSC for bone [22] to describe the overlap between the CT and synthetic CT bone volumes.

When we compared the two loss functions (MAE and MI), we found that using MI better compensates for misalignments between MRI and CT images. As with atlas-based models [12], accurately aligning MRI/CT images is critical for convolutional neural network (CNN) models. Misalignments in training data may introduce inaccuracies into the model that ultimately lead to inaccurate predictions by the trained model [23]. In general, when patients transfer from one imaging device like MRI to another imaging device like CT, some anatomical changes can happen. This training data set is called unregistered, and it needs to be accurately aligned. However, it is difficult to perfectly register the MRI/CT images due to different table sizes and organ deformation, especially in the pelvic domain. To overcome this problem, we used the MI as the loss function. MI quantifies the "amount of information" of one variable when another variable is known. Maximizing Mutual Information is equivalent to minimizing the joint entropy (joint histogram). By including joint entropy in the loss function, we calculated the amount of information in the output slice (generated image) based on the ground truth slice (real image). Moreover, a gradient descent optimizer (e.g. Adam) updates weights (model's parameters) in the direction of minimizing joint entropy via backward propagation. The loss function (joint histogram) value is low when images are aligned and high when images are not aligned. So, the misalignments between two images were implicitly fixed in this manner. This concept has been proven in the registration framework using Mutual Information as the loss function [24, 25].

**Table 2:** The mean ± SD of Gamma pass rates for 2 criteria: $\gamma_{1\%/1mm}$, and $\gamma_{2\%/2mm}$ calculated over 14 test patients (using the same plan for synthetic CT images).

| Case number | 3D Gamma passing rate (%) | | 2D Gamma passing rate (%) | |
| --- | --- | --- | --- | --- |
| | 2%/2mm | 1%/1mm | 2%/2mm | 1%/1mm |
| 1 | 99.9 | 97.8 | 99.7 | 95.6 |
| 2 | 97.2 | 92.5 | 96.1 | 91.9 |
| 3 | 99.7 | 96.7 | 99.2 | 99.1 |
| 4 | 98.8 | 94.5 | 97.8 | 88.5 |
| 5 | 99.2 | 95.1 | 98.6 | 94.9 |
| 6 | 98.2 | 89.8 | 97.9 | 91.4 |
| 7 | 99.8 | 98.5 | 99.8 | 97.7 |
| 8 | 99.6 | 93.4 | 98.7 | 89.9 |
| 9 | 99.7 | 96.5 | 99.7 | 96.4 |
| 10 | 99.3 | 97.9 | 99.9 | 95.4 |
| 11 | 99.8 | 89.6 | 99.6 | 90.2 |
| 12 | 99.3 | 96.2 | 97.9 | 97.1 |
| 13 | 98.6 | 94.8 | 97.8 | 93.4 |
| 14 | 99.2 | 91.6 | 98.3 | 89.6 |
| **Mean** | **99.2** | **94.6** | **98.7** | **93.6** |
| **SD** | **0.8** | **2.9** | **1.1** | **3.4** |

Importantly, electron density maps for dose calculation and geometric accuracy are essential in MRI-only treatment planning platforms. Gamma analysis of the dose recalculated for the same treatment plans indicated that the synthetic CT images achieved higher passing rates than the typical 95% passing rates for the 3%/3 mm gamma criteria used for treatment delivery quality assurance. DVH parameters were also equivalent to real CT for both target and normal structures. Target prescription coverage was within 1%, and calculated doses to brainstem, chiasm, and optic nerve were within 1% at the clinically critical 54 Gy dose level. The dosimetric results showed that the accuracy of the generated synthetic CT images was sufficient to produce clinically equivalent treatment plans.

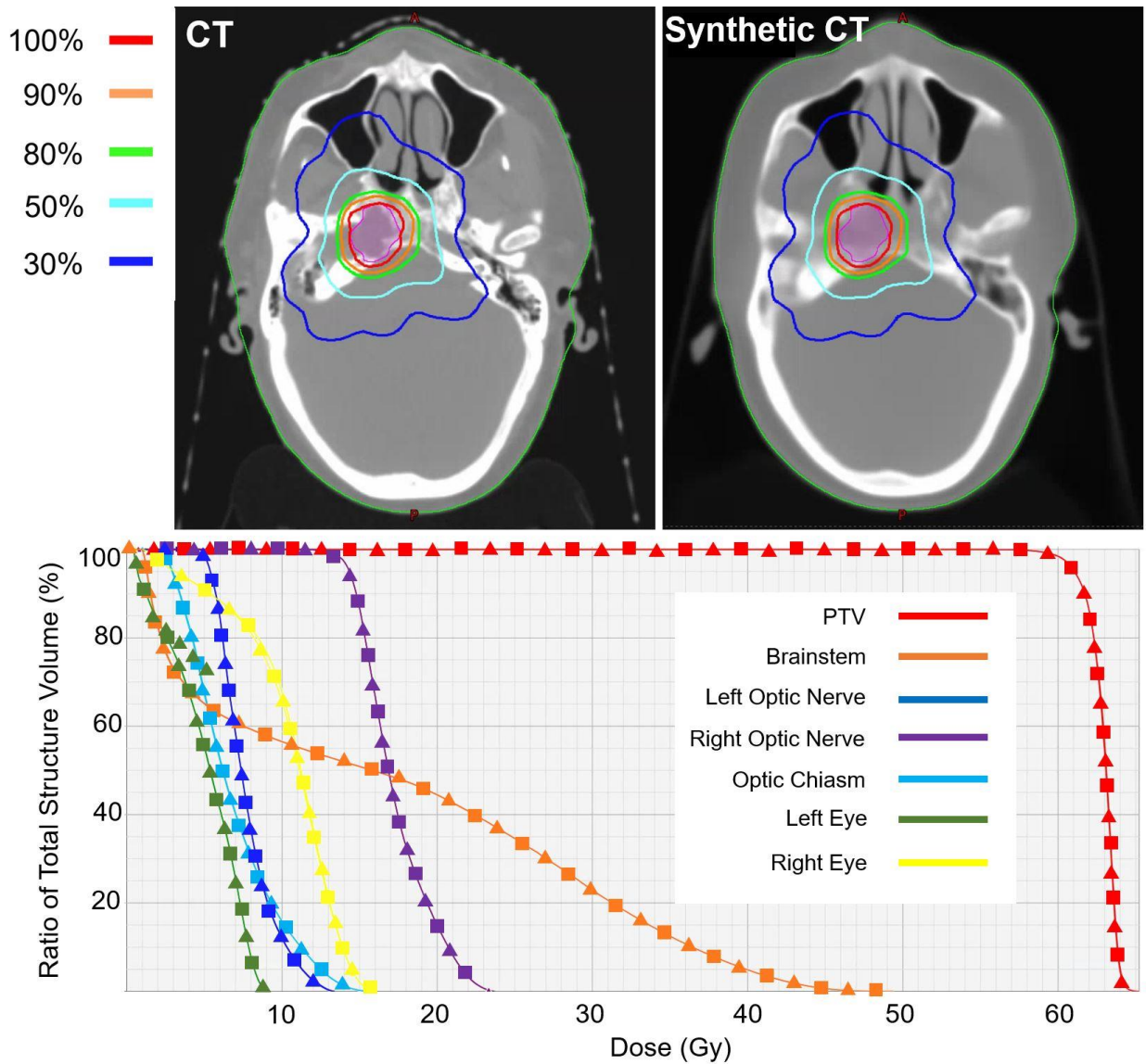

**Figure 4.** (A) Dose comparison between CT and synthetic CT for a single VMAT plan. (B) DVH plot with corresponding PTV and OARs from one patient. Squares and triangles represent mean DVH values from CT and synthetic CT, respectively.

The mean HU MAE values between synthetic CT and real CT that we calculated in this study were lower than those observed in current published methods. Higher MAE values have been achieved with atlas-based methods [26, 27]. Registration errors between real CT and MRI may contribute to an incorrect HU value and targeting during rigid and deformable registrations. For example, Sjolund *et al.* [28] reported a mean MAE of 123.2 ± 18.6 HU in brain MRI images from 10 patients using atlas-based registration and a fusion strategy. Dinkla *et al.*[29] applied a dilated CNN to $T_1$- weighted images to generate synthetic CT images and achieved an MAE of 67 ± 11 HU. A recent study by Emami *et al.*[30] using GAN achieved an MAE of 89.3 ± 10.3 HU. Several groups provided different promising approaches for synthetic CT generation, which were summarized in three recent review papers [22, 31, 32]. Comparisons between existing methods using performance metrics are highly dependent on HU value, FOV, and resolution.

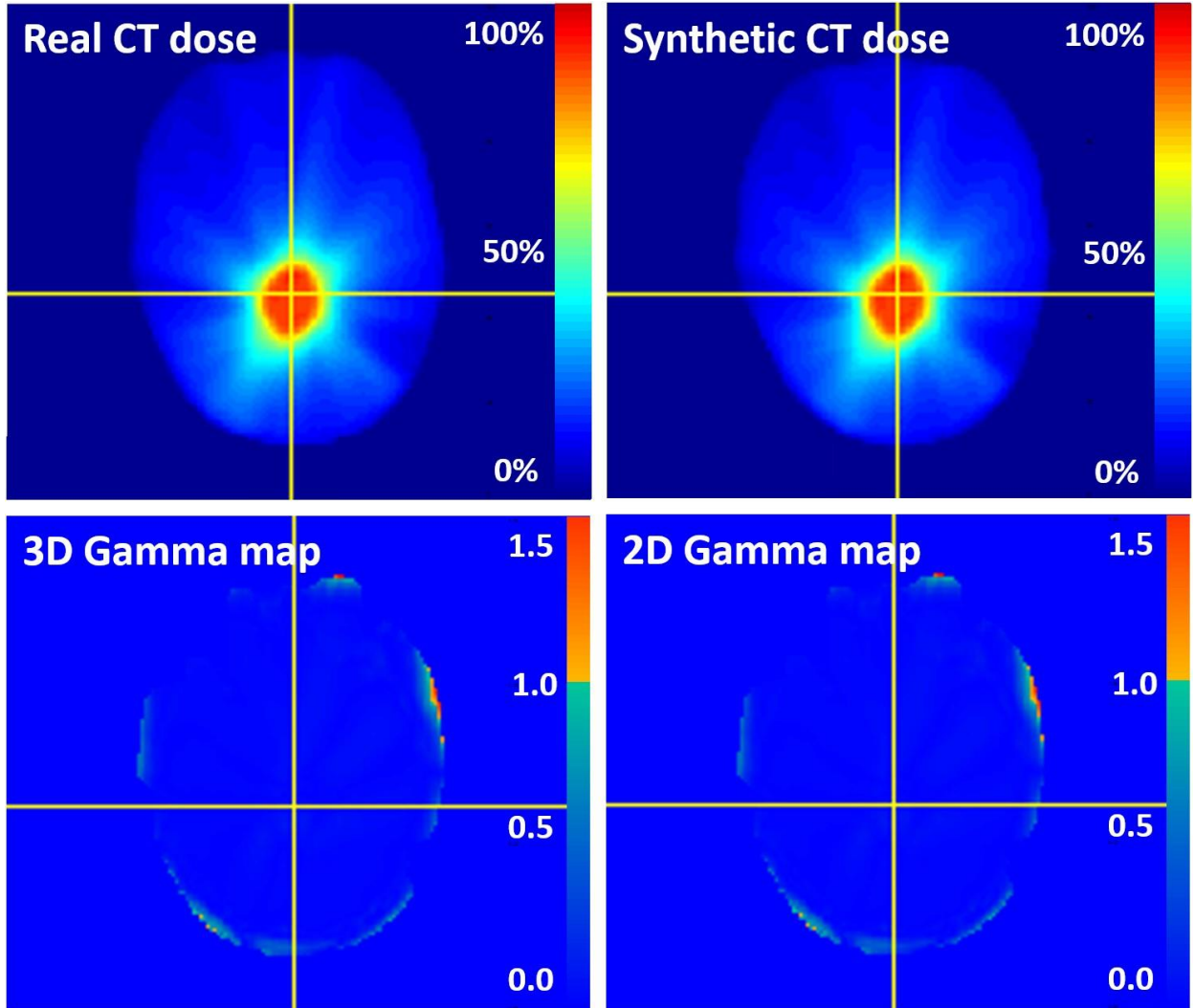

**Figure 5.** Top row images show the dose distribution map. Bottom row images show the 2D and 3D gamma analysis comparison between CT and synthetic CT.

Our proposed method provides accurate and reproducible synthetic CT images using a GAN model, which can increase efficiency, accuracy, and precision in a clinical workflow. The computational time in our study was about 1 second per patient for generating synthetic CT images, and training time was 33 hours using an NVIDIA Tesla K80 dual-GPU graphic card. The MI loss function allows the model to use unregistered data to generate synthetic CT images. This will eliminate the need to preregister MRI and CT images, and it will streamline the process of using MRI images from radiology, which suffice for generating synthetic CT images. Importantly, this model accurately distinguishes between air and bone regions, which is a major challenge in generating synthetic CT images and is the most substantial contribution to dose calculation error. Future work is warranted to develop 3D GAN models, using larger training datasets, for generating synthetic CT images to be used in MRI-only radiotherapy.

In conclusion, MRI-only treatment planning will reduce radiation dose, patient time, and imaging costs associated with CT imaging, streamlining clinical efficiency and allowing high-precision radiation treatment planning. Despite these advantages, several challenges prevent clinical implementation of MRI-only radiation therapy. Through the method we have proposed here, synthetic CT images can be generated from only one pulse sequence of MRI images of a range of brain tumors. This method is a step toward using artificial intelligence to establish MRI-only radiation therapy in the clinic.